\documentclass[aps,pra,showpacs,twocolumn]{revtex4}
\usepackage{graphicx}
\usepackage{bm}
\usepackage{amssymb}

\newcommand{\be}{\begin{equation}}
\newcommand{\ee}{\end{equation}}
\newcommand{\bea}{\begin{eqnarray}}
\newcommand{\eea}{\end{eqnarray}}
\newcommand{\beas}{\begin{eqnarray*}}
\newcommand{\eeas}{\end{eqnarray*}}

\begin{document}
\title{Scaling of success probabilities for linear optics gates}
\author{Stefan Scheel}
\email{s.scheel@imperial.ac.uk}
\author{Koenraad M.R.~Audenaert}
\email{k.audenaert@imperial.ac.uk}
\affiliation{Quantum Optics and Laser Science, Blackett Laboratory,
Imperial College London, Prince Consort Road, London SW7 2BW, UK}

\begin{abstract}
By using the abstract linear-optical network derived in [S.~Scheel and
N.~L\"utkenhaus, New J. Phys. \textbf{6}, 51 (2004)] we show
that for the lowest possible ancilla photon numbers the
probability of success of realizing a (single-shot) generalized
nonlinear sign shift gate on an ($N+1$)-dimensional signal state
scales as $1/N^2$. We limit ourselves to single-shot gates without
conditional feed-forward. We derive our results by using
determinants of Vandermonde-type over a polynomial basis which is
closely related to the well-known Jacobi polynomials.
\end{abstract}

\date{\today}

\pacs{03.67.-a,42.50.-p,42.50.Ct}
\maketitle

\section{Introduction}

One of the promising routes to implementing small-scale quantum
information processing networks for tasks in quantum communication or
cryptography is by using linear-optical networks. Already a number
of experiments have been performed that accomplish the task of
generating controlled-NOT gates as the basic nontrivial
two-mode operation \cite{Pittman,White}. The necessary nonlinearities
on the single-photon level are produced via conditional measurements
\cite{Ban94,Clausen99}.

Some two-mode gates such as the controlled-$\sigma_z$ gate
can be generated by acting separately on both modes within a
Mach--Zehnder interferometric setup. In that way the complexity of a
two-mode gate is reduced to a single-mode gate, the nonlinear sign
shift gate. A number of recent theoretical works
\cite{KLM,Ralph01,Scheel03,Lapaire03} give a variety of networks
capable of performing the nonlinear sign shift gate with the following
transformation rule:
\begin{equation}
c_0|0\rangle +c_1|1\rangle +c_2|2\rangle \mapsto
c_0|0\rangle +c_1|1\rangle -c_2|2\rangle
\end{equation}
which can serve as a building block for the two-mode controlled-phase
gate. The crucial question for the ability to concatenate
linear-optical gates are their probabilities of success. Recently, it
has been shown that the nonlinear sign shift gate cannot be realized
with a success probability of more than $1/4$ if one abstains from
using conditional feed-forward \cite{Scheel04b,Eisert} which is
stronger than the bound of $1/2$ obtained in
\cite{Knill03}.

\begin{figure}[t]
\centerline{\includegraphics[width=8cm]{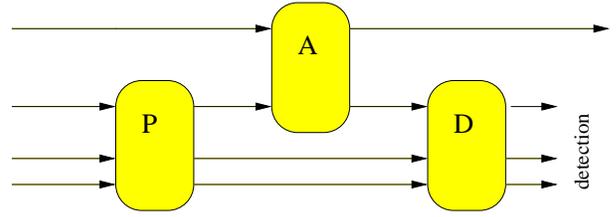}}
\caption{\label{fig:decomposition} Preparation stage `P' of the
ancilla state that is fed into the `active' beam splitter `A', and
decomposition stage `D' for detection.}
\end{figure}

These results were obtained by considering an abstract type of network
that includes all possible networks by dividing it into an `active'
beam splitter `A' that mixes signal and ancilla states, a preparation
stage `P' and a detection stage `D' that are there to generate and
disentangle a purpose-specific ancilla, respectively. The general
scheme is reproduced in Fig.~\ref{fig:decomposition}.
In short, the abstraction consists in the observation that each U($N$)
network can be decomposed into a triangle-shaped network of at most
$N(N-1)/2$ beam splitters (and some additional phase shifters)
\cite{Reck94} which in fact is nothing else but doing a
Householder-type diagonalization of the unitary matrix associated with
the network. Then, it is easy to see that the signal mode can be
chosen to impinge only on a single beam splitter (here called `A')
with the rest of the network conveniently divided into the ancilla
preparation and detection stages `P' and `D', respectively.

What we will try to do in this article is to  answer the question how
the probability of success scales as the signal state becomes
higher-dimensional, thereby giving a hint on the possible scaling law
for multi-mode quantum gates (by virtue of the above-mentioned fact
that generalized nonlinear sign shift gates can be combined in
Mach--Zehnder interferometer-type set-ups to yield multi-mode
controlled-phase gates). However, as we have seen previously, a
general answer is hard to obtain. Fortunately, it turned out that the
most important special case can be dealt with analytically. It has
been argued already in \cite{Scheel04b} and confirmed in \cite{Eisert}
that the success probability, taken as a function of the chosen
ancilla photon numbers, reaches its maximum if the ancilla contains as
few photons as possible. In
addition, the restriction to the lowest possible photon number is
justified by the fact that in this case the least quantum-state
engineering is needed to generate the ancilla in the preparation stage
`P' and decoherence affects the ancilla only minimally.

This article is organized as follows. In Sec.~\ref{sec:defs} we
provide the necessary framework that is suitable for our set task.
Specializing to the lowest admissible photon numbers, we show in
Sec.~\ref{sec:proof} that there exist networks that realize the
generalized nonlinear sign  shift (to be defined in
Sec.~\ref{sec:defs}) with a probability of success of $1/N^2$ which
constitutes the main result of our paper. Some details of the
calculation are provided in the Appendices. Conclusions are drawn
in Sec.~\ref{sec:conclusions}.

\section{($N+1$)-dimensional signal state with $N$-dimensional ancilla}
\label{sec:defs}

A possible building block for multi-mode quantum gates via some
Mach--Zehnder interferometer-type set-up \cite{Scheel03} is the
generalized nonlinear sign shift gate. Let us therefore consider an
($N+1$)-dimensional signal state
\begin{equation}
|\psi_{\text{in}}\rangle = \sum\limits_{k=0}^N c_k |k\rangle \,.
\end{equation}
The generalized nonlinear sign shift gate is defined by the
following transformation rule:
\begin{equation}
c_N \mapsto -c_N \,,
\end{equation}
leaving all other coefficients untouched. As thoroughly described in
\cite{Scheel04b}, not all ancilla states are admissible. In fact, the
ancilla must consist of a superposition of states with fixed photon
number. Moreover, the number of photons that are detected after the
disentangling stage `D' in Fig.~\ref{fig:decomposition} must be the
same as the initial ancilla photon number. The reason for this
requirement is essentially that the Hilbert spaces of incoming and
outgoing signal states have to be isomorphic.

We will restrict ourselves to the case in which we use an
$N$-dimensional ancilla for transforming an ($N+1$)-dimensional signal
state. This choice is somewhat restrictive but, in view of minimizing
experimental resources, the most accessible choice.
Let us denote the $N$ different photon numbers in the ancilla by
$n_l$, $l=1\ldots N$. Then, the ancilla can be represented in the form
\begin{equation}
\sum\limits_{l=1}^N \gamma_l |n_l\rangle |A_{n-n_l}\rangle
\end{equation}
with $\sum_l\gamma_l^2=1$ (the expansion coefficients can be taken to
be real) and $n=\sum_l n_l$. The state $|A_{n-n_l}\rangle$ is
arbitrary as concerns the number of modes as long as it contains
exactly $n-n_l$ photons. The advantage of introducing the abstract
network in Fig.~\ref{fig:decomposition} is apparent since the
$|A_{n-n_l}\rangle$ do not play any role in what follows.

Now we transform the signal and ancilla states at the `active' beam
splitter `A' and use the fact that the matrix elements
$\langle k,n_l|\hat{U}|k,n_l\rangle$ can be written as
\begin{equation}
\langle k,n_l|\hat{U}|k,n_l\rangle = (T^\ast)^{n_l-k}
P_k^{(0,n_l-k)}(2|T|^2-1) \,,
\end{equation}
where $P_k^{(0,n_l-k)}(x)$ is a Jacobi polynomial defined by
\begin{equation}
\label{eq:jacobi}
P_k^{(0,n_l-k)}(2|T|^2-1) = \sum\limits_{m=0}^k {k \choose m}
{n_l+m \choose m} (|T|^2-1)^m \,.
\end{equation}
and $T$ the (in general complex) transmission coefficient of the beam
splitter `A'. Then, the (unnormalized) states $|\psi_k\rangle$
($k=1\ldots N+1$) after the transformation can be defined as
\begin{equation}
|\psi_k\rangle = \sum\limits_{l=1}^N \gamma_l
(T^\ast)^{n_l-k+1} P_{k-1}^{(0,n_l-k+1)}(P)
|n_l\rangle|A_{n-n_l}\rangle \,,
\end{equation}
where we have abbreviated the argument $P=2|T|^2-1$ of the Jacobi
polynomial to emphasize its dependence on the permanent of the beam
splitter matrix. This is a reminder of the fact that matrix elements of
unitary operators can always be written as permanents of a matrix
associated with the symmetric tensor power of the beam splitter matrix
\cite{Scheel04c}.

The state we project upon is denoted by
\begin{equation}
|\psi\rangle = \sum\limits_{l=1}^N \alpha_l^\ast |n_l\rangle
|A_{n-n_l}\rangle
\end{equation}
where $\sum_l|\alpha_l|^2=1$. Then, in order to realize the generalized
nonlinear sign shift, we need to fulfil the $N$ equations
\begin{equation}
\label{eq:ndim1}
\langle\psi|\psi_{N+1}\rangle = -
\langle\psi|\psi_k\rangle\,,\quad k=1\ldots N \,.
\end{equation}
That specifies all weights $\alpha_l$ as well as the transmission
coefficient $T$ which can be used to compute the probability as
\begin{equation}
\label{eq:ndim3}
p= |\langle\psi|\psi_k\rangle|^2 \quad \mbox{for any }k\,.
\end{equation}
In fact, Eq.~(\ref{eq:ndim1}) is a homogeneous matrix equation of the
form
\begin{equation}
\label{eq:ndim1a}
\sum\limits_{l=1}^N a_{kl} \gamma_l \alpha_l = 0
\end{equation}
with matrix elements
\begin{eqnarray}
\label{eq:ndimcoeff}
a_{kl} &=& a_{kl}^{(1)} + a_{kl}^{(2)} \nonumber \\
&=& (T^\ast)^{n_l-N} P_N^{(0,n_l-N)}(P) \nonumber \\
&& + (T^\ast)^{n_l-k+1} P_{k-1}^{(0,n_l-k+1)}(P) \,.
\end{eqnarray}
From linear algebra it is known that the homogeneous system of
equations (\ref{eq:ndim1a}) has nontrivial solutions only if the
determinant of the coefficient matrix vanishes. That gives a condition
for the transmission coefficient $T$.

From the expansion theorem of determinants we immediately have the
solutions for $\alpha_l\gamma_l$ given in terms of the cofactors
$A_{kl}$ of the matrix elements $a_{kl}$,
i.e. $\alpha_l\gamma_l\propto A_{kl}$. Because the success probability
(\ref{eq:ndim3}) is nothing else but
$p=|\sum_l\alpha_l\gamma_la_{kl}^{(2)}|^2$, it is clearly maximized if
the magnitudes of $\alpha_l$ and $\gamma_l$ are the same. From the
normalization condition for $|\psi\rangle$ we then get the solution
\begin{equation}
\label{eq:prob}
p_{\text{max}} = \frac{\left| \sum_l A_{kl}a_{kl}^{(2)}  \right|^2}
{(\sum_l |A_{kl}|)^2} \quad \mbox{for some }k\,.
\end{equation}
We will shortly see that the choice of $k$ does not matter.

\section{Maximal success probability}
\label{sec:proof}

In most cases, one needs to refer to numerical methods to maximize the
success probability (\ref{eq:prob}) for a given set of ancilla photon
numbers $n_l$. However, there is one case in which an analytical
result can be obtained.
For this purpose, let us consider the photon numbers $n_l=l-1$. This
is the ancilla with the lowest possible photon numbers and as such the
experimentally most important situation. Then, we find that there
exists at least one solution of the form
\begin{equation}
T = 1 - 2^{1/N} \,.
\end{equation}
For $N=2$, i.e. the nonlinear sign shift gate with up to two photons
in the signal state, this is the solution $T=1-\sqrt{2}$ encountered
in \cite{KLM,Scheel04b}. For this particular choice, the maximal
success probability reaches
\begin{equation}
\label{eq:scale}
p_{\text{max}} = \frac{1}{N^2} \,.
\end{equation}
This result is obviously true for $N=1$ since the corresponding gate
can indeed be done deterministically, and for $N=2$ that is the result
obtained in \cite{Scheel04b,Eisert}. Note the surprising fact that the success
probability does not scale exponentially but rather quadratically with
$N$. Equation~(\ref{eq:scale}) constitutes the main result of this
paper. We will spend the following subsections deriving it.

\subsection{Determinant of the coefficient matrix}
\label{sec:det}

In the calculation of the projection vector $|\psi\rangle$ we
encountered the matrix consisting of the elements $a_{kl}$
[Eq.~(\ref{eq:ndimcoeff})]. As one can see, the $a_{kl}^{(1)}$ are
actually independent of the row index $k$ and therefore constant
columns. In addition, each column consists of two sub-columns, and the
determinant can thus be expanded as a sum of determinants, each of
which contains one of the sub-columns \cite{VeinDale}. Hence,
\begin{equation}
\det (a_{kl}) = \sum\limits_{m_1,\ldots,m_N=1}^2 \det (a_{kl}^{(m_l)}) \,.
\end{equation}
Together with the property that a determinant vanishes if it contains
more than one constant column, this means that
only $N+1$ out of the total $2^N$ terms eventually have to be computed.
The first term is just $\det (a_{kl}^{(2)})$, and the other
$N$ terms are obtained by replacing the $m$th column of
$(a_{kl}^{(2)})$ by the corresponding column of $(a_{kl}^{(1)})$. Thus,
\begin{equation}
\label{eq:det}
\det (a_{kl}) = \det (a_{kl}^{(2)}) +\sum\limits_{m=1}^N
\det (a_{kl}^{(2)}: a_{km}^{(2)}\mapsto a_{km}^{(1)}) \,.
\end{equation}
All higher-order replacements vanish due to the fact that determinants
of matrices with more than one constant column vanish identically.

Let us first consider the term $\det (a_{kl}^{(2)})$. By inspection of
the structure of the matrix elements $a_{kl}^{(2)}$ one realizes that
this determinant can be written as
\begin{equation}
\label{eq:detakl2}
\det (a_{kl}^{(2)}) =
\det \left[P_{k-1}^{(0,l-k)}(2|T|^2-1)\right] \,.
\end{equation}
Due to the special choice of photon number $n_l=l-1$, all prefactors
containing the transmission coefficient cancel out. The next
observation is crucial for all the following. The Jacobi polynomials
$P_{k-1}^{(0,l-k)}(2|T|^2-1)$, being polynomials of order $k-1$ in
$2|T|^2-1$, are also polynomials of order $k-1$ in the photon numbers
$n_l=l-1$. This can be seen as follows. Recall the definition of the
Jacobi polynomials [Eq.~(\ref{eq:jacobi}), we briefly revert to the
general notation $n_l$ for the photon numbers] and rewrite
the binomial factor ${n_l+m \choose m}$ as
\begin{equation}
\label{eq:sigma}
{n_l+m \choose m} = \frac{1}{m!} \prod\limits_{i=1}^m (n_l+i) =
\frac{1}{m!} \sum\limits_{p=0}^m n_l^p \sigma_{m-p}^{(m)}
\end{equation}
which itself is a polynomial of degree $m$ in $n_l$. Hence, the Jacobi
polynomial $P_k^{(0,n_l-k)}(2|T|^2-1)$ is also a polynomial of degree
$k$ in the photon numbers $n_l$. The $\sigma_{m-p}^{(m)}$ are the
elementary symmetric polynomials of degree $m-p$ of $m$ variables
which in our case are just the integers $1\ldots m$. We will now
define a new class of polynomials $S_k^{(x)}(n)$ by setting
\begin{equation}
\label{eq:spoly1}
S_k^{(x)}(n_l) = \sum\limits_{p=0}^k c_{kp} n_l^p
= \sum\limits_{p=0}^k n_l^p \sum\limits_{m=p}^k
{k \choose m} \frac{(x^2-1)^m}{m!} \sigma_{m-p}^{(m)} \,.
\end{equation}
These polynomials fulfil the identity
\begin{equation}
S_k^{(|T|)}(n_l) \equiv P_k^{(0,n_l-k)}(2|T|^2-1) \,.
\end{equation}
From their definition, Eq.~(\ref{eq:spoly1}), it is easy to show that
these polynomials fulfil a three-term recursion relation. Some
properties of the $S$-polynomials are collected in Appendix~\ref{sec:appa}.

With this definition, we can finally rewrite the determinant
(\ref{eq:detakl2}) with the help of the $S$-polynomials as
\begin{equation}
\det (a_{kl}^{(2)}) =
\det \left[S_{k-1}^{(|T|)}(n_l)\right] \,,
\end{equation}
which is nothing but a Vandermonde determinant over the $S$-polynomial
basis. For later use, let us denote this Vandermondian by
$V_{S,N}(\{n_l\})$, where the subscripts $S$ and $N$ denote the
polynomial basis and the dimension, respectively.
For such a determinant we have the relation (see, e.g. \cite{VeinDale})
\begin{equation}
V_{S,N}(\{n_l\})=V_N(\{n_l\}) \prod\limits_{k=0}^{N-1} c_{kk}
\end{equation}
where $V_N(\{n_l\}$ is the usual Vandermonde determinant of the photon
numbers $n_l$. This implies that
\begin{equation}
\label{eq:detsn}
\det (a_{kl}^{(2)}) = \left( \prod\limits_{l=1}^N \frac{1}{(l-1)!}
\right) (|T|^2-1)^{\frac{N}{2}(N-1)} V_N(\{n_l\}) \,.
\end{equation}
Specializing again to $n_l=l-1$, in which case
$V_N(\{n_l\})=\prod_l (l-1)!$, we are left with
\begin{equation}
\label{eq:simpledet}
\det (a_{kl}^{(2)}) =
(|T|^2-1)^{\frac{N}{2}(N-1)} \,.
\end{equation}

Next, in order to compute the determinant in Eq.~(\ref{eq:det}) we
also need the terms in which the $m$th column of $(a_{kl}^{(2)})$ has
been replaced by $(a_{km}^{(1)})$. Expanding the determinants
$\det (a_{kl}^{(2)}: a_{km}^{(2)} \mapsto a_{km}^{(1)})$ with respect
to the column $m$, we obtain
\begin{eqnarray}
\label{eq:detsum1}
\sum\limits_{m=1}^N
\det (a_{kl}^{(2)}: a_{km}^{(2)} \mapsto a_{km}^{(1)})
=\sum\limits_{m,n=1}^N a_{nm}^{(1)} A_{nm}^{(2)} \,,
\end{eqnarray}
where the $A_{nm}^{(2)}$ are the cofactors of the matrix
$(a_{kl}^{(2)})$. They are, disregarding some prefactors, nothing
but generalized Vandermondians over the $S$-polynomial basis with one
gap, denoted by $V_{S,N\backslash \{n-1\}}^{(m-1)}$, with the
polynomials of order $n-1$ and photon number $n_m=m-1$ missing.
With that, Eq.~(\ref{eq:detsum1}) becomes
\begin{eqnarray}
\label{eq:detsum2}
\lefteqn{
\sum\limits_{m=1}^N
\det (a_{kl}^{(2)}: a_{km}^{(2)} \mapsto a_{km}^{(1)}) }
\nonumber \\ && =
\sum\limits_{n=1}^N (-1)^{N-n} (T^\ast)^{n-N-1}
\nonumber \\ &&  \times
\sum\limits_{m=1}^N (-1)^m  S_N^{(|T|)}(m-1)
V_{S,N\backslash \{n-1\}}^{(m-1)} \,.
\end{eqnarray}
The last line in Eq.~(\ref{eq:detsum2}) is just the definition of
$V_{S,N+1\backslash \{n-1\}}$. Such a quantity can be computed in the
following way. Analogously to the elementary symmetric polynomials
$\sigma_i^{(j)}$ defined in Eq.~(\ref{eq:sigma}) we define the
elementary symmetric polynomials $s_i(x;N;N)$ with respect to the
$S$-polynomial basis by
\begin{equation}
\label{eq:defsymms}
\frac{(x^2-1)^N}{N!} \prod\limits_{i=1}^N (n-n_i) =
\sum\limits_{j=0}^N (-1)^{N-j} s_{N-j}(x;N;N) S_j^{(x)}(n) \,.
\end{equation}
These polynomials $s_i(x;N;N)$ are $i$th-order symmetric
polynomials over the $N$ different photon numbers $n_l$,
$l=1\ldots N$. Expanding $V_{S,N+1}$ on one hand with respect to a
Vandermondian of lower order and on the other hand with respect to its
last column, viz.
\begin{eqnarray}
V_{S,N+1} &=& \frac{(x^2-1)^N}{N!} \prod\limits_{i=1}^N (n_{N+1}-n_i)
V_{S,N} \nonumber \\ &=& \sum\limits_{m=0}^N (-1)^{N-m}
S_m^{(x)}(n_{N+1}) V_{S,N+1\backslash \{m\}}
\end{eqnarray}
and using the definition (\ref{eq:defsymms}) for the symmetric
polynomials, we obtain that
\begin{equation}
V_{S,N+1\backslash \{n-1\}} = s_{N-n+1}(|T|;N;N) V_{S,N}
\end{equation}
(see also \cite{Werther93}).

The symmetric polynomials $s_i(x;N;N)$ can be computed
by inserting Eq.~(\ref{eq:sigma}) into Eq.~(\ref{eq:defsymms}).
When specifying yet again to the choice of photon numbers $n_l=l-1$,
the solution is simply
\begin{equation}
\label{eq:spoly2}
s_k(|T|;N;N) = {N \choose k} |T|^{2k}
\end{equation}
which can easily be checked by inserting into Eq.~(\ref{eq:defsymms})
and using the definition (\ref{eq:spoly1}) for the
$S$-polynomials.
Note that the symmetry with respect to the
interchange of any two photon numbers $n_k=k-1$ and  $n_l=l-1$ is
hidden inside the binomial coefficient in Eq.~(\ref{eq:spoly2}).
For a constructive proof of the statement (\ref{eq:spoly2}) and
generalizations of it, we refer the reader to Appendix~\ref{sec:appb}.

Combining the knowledge about the symmetric polynomials and the
Vandermondians defined earlier, we arrive at the expression
\begin{eqnarray}
\lefteqn{
\sum\limits_{n,m=1}^N (-1)^{N-n+m} (T^\ast)^{n-N-1} S_N^{|T|}(m-1)
V_{S,N\backslash\{n-1\}}^{(m-1)}
} \nonumber \\ && 
=\sum\limits_{n=1}^N (-1)^{N-n} (T^\ast)^{n-N-1}
V_{S,N+1\backslash \{n-1\}} \nonumber \\ &&
=\sum\limits_{n=1}^N (-1)^{N-n} (T^\ast)^{n-N-1} {N \choose n-1}
|T|^{2(N-n+1)} V_{S,N} \nonumber \\ &&
=(|T|^2-1)^{\frac{N(N-1)}{2}} \left[ 1-(1-T)^N \right] \,.
\end{eqnarray}
Combined with the result for the determinant $\det(a_{kl}^{(2)})$
[Eq.~(\ref{eq:simpledet})] this yields the final result for the
determinant of the coefficient matrix as
\begin{equation}\label{eq:deta}
\det(a_{kl}) =
(|T|^2-1)^{\frac{N(N-1)}{2}} \left[ 2-(1-T)^N \right] \,.
\end{equation}
A nontrivial condition for this determinant to vanish is thus
\begin{equation}
\label{eq:solution}
T=1-2^{1/N}
\end{equation}
just as we set out to show. What we have seen is that calculating the
value of the transmission coefficient $T$ of the active beam splitter
`A' is still mathematically involved despite the reduction of the
problem to the abstract network.

\subsection{Success probability}

In order to compute the success probability (\ref{eq:prob}), we
need to calculate the cofactors $A_{kl}$ of the generalized
Vandermondian $(a_{kl})$. It turns out that it is useful to choose
$k=N$ in this relation since the cofactors $A_{Nl}$ are essentially
generalized Vandermondians over the $S$-polynomial basis which can be
computed in the same way as shown above. Although this particular
choice for $k$ is not unique, it represents the most easily
tractable situation. Clearly, any other $k$ could have been chosen.
However, since the solution $T=1-2^{1/N}$ is unique and is a simple
root of Eq.~(\ref{eq:ndim1a}), the result cannot depend on the
particular choice of $k$. Since now $T\in \mathbb{R}$, we can drop
asterisks to simplify notation.

The numerator in Eq.~(\ref{eq:prob}) can be calculated in the
following way. By using simple row manipulations, it is shown to be
equal to
\begin{eqnarray}
\sum\limits_{l=1}^N a_{Nl}^{(2)} A_{Nl} &=& \det (a_{kl}^{(2)})
+\sum\limits_{n=1}^{N-1} \det (a_{kl}^{(2)}:a_{nl}^{(2)}\mapsto
a_{nl}^{(1)}) \nonumber \\ &=&
-\det (a_{kl}^{(2)}:a_{Nl}^{(2)}\mapsto a_{Nl}^{(1)}) \,,
\end{eqnarray}
where we have made use of the fact that the determinant $\det(a_{kl})$
vanishes for $T=1-2^{1/N}$. All subsequent formulas have to be
understood with the solution for $T$ in mind.

Inserting the definitions for the matrix elements $a_{kl}^{(1,2)}$, it
turns out that this is a Vandermondian with one gap. More precisely,
\begin{eqnarray}
\lefteqn{
\det (a_{kl}^{(2)}:a_{Nl}^{(2)}\mapsto a_{Nl}^{(1)}) = } \nonumber \\ &&
-T^{-1} V^{(N)}_{S,N+1\backslash\{N-1\}} \,.
\end{eqnarray}
Expanding this determinant with respect to the last row, we get
\begin{equation}
V^{(N)}_{S,N+1\backslash\{N-1\}} = \sum\limits_{k=0}^{N-1} (-1)^{N+k+1}
S_N^{(T)}(k) V_{S,N-1}^{(k)} \,.
\end{equation}
The Vandermondian over the polynomial basis can easily be computed
from the corresponding Vandermondian over the power basis as (see
Appendix~\ref{sec:appc})
\begin{equation}
\label{eq:onegap}
V_{S,N-1}^{(k)} =
(T^2-1)^{\frac{N-1}{2}(N-2)} {N-1 \choose k}\,.
\end{equation}
With that and the definition of the $S$-polynomials one finds that
\begin{eqnarray}
\lefteqn{
\sum\limits_{k=0}^{N-1} (-1)^{N+k+1} {N-1 \choose k} S_N^{(T)}(k) =}
\nonumber \\ && \hspace*{-2ex}
\sum\limits_{k=0}^{N-1} \sum\limits_{m=1}^N (-1)^{N+k+1}
{N-1 \choose k} {k+m \choose m} {N \choose m} (T^2-1)^m
\nonumber \\ && =
\sum\limits_{m=1}^N {N \choose m} (T^2-1)^m {m \choose N-1}
\nonumber \\ && = NT^2(T^2-1)^{N-1}
\end{eqnarray}
where the last line follows from the fact that only $m=N-1,N$ give
non-zero contributions to the sum. Combining all the results, we find
that
\begin{equation}
\label{eq:num}
\sum\limits_{l=1}^N a_{Nl}^{(2)} A_{Nl} =
NT(|T|^2-1)^{\frac{N}{2}(N-1)} \,.
\end{equation}


The remaining task is to compute the normalization factor
$(\sum_l |A_{Nl}|)^2$ for $T=1-2^{1/N}$.
Recalling the expression (\ref{eq:ndimcoeff}) for the matrix elements
$a_{kl}$, inserting the expression
(\ref{eq:jacobi}) for the Jacobi polynomials, and changing the
indexation such that $k$ and $l$ now range from 0 to $N-1$, we get
\be
a_{kl} = \sum_{m=0}^N {l+m\choose m}(T^2-1)^m T^l
\left[{N\choose m}T^{-N}+{k\choose m}T^{-k}\right].
\ee
To find the cofactors $A_{kl}$ of $a=(a_{kl})_{0\le k,l\le N-1}$, we
make use of the fact that they are given by
$$
A_{kl} = \det(a) (a^{-1})_{lk}.
$$
We need $A_{N-1,l}$, and this can thus be found from solving the
equation $ax=b$, where $b_k=\delta_{k,N-1}$, and
$A_{N-1,l} = \det(a) x_l$.
The equation $ax=b$ gives rise to the following $N$ equations in $x$:
\be\label{eq:system}
\sum_{m=0}^N \left[{N\choose m}T^{-N}+{k\choose m}T^{-k}\right] y_m =
\delta_{k,N-1}\,,
\ee
where we have defined the $N+1$ quantities $y_m$ as
\be\label{eq:ym}
y_m = (T^2-1)^m \sum_{l=0}^{N-1} {l+m\choose m} T^l x_l \,.
\ee
We will now proceed by inverting the two systems (\ref{eq:system}) and
(\ref{eq:ym}) in succession.
The former system is of course underdetermined, which means that the
solution for $y_m$ will contain one free parameter.
This parameter will be solved for when solving (\ref{eq:ym}) for $x_l$.
Let us now choose as free parameter $\eta$ with
\be\label{eq:eta}
\eta = \sum_{m=0}^N {N\choose m}T^{-N} y_m,
\ee
then (\ref{eq:system}) turns into
\be
\sum_{m=0}^N {k\choose m} y_m = (\delta_{k,N-1}-\eta)T^k.
\ee
Since $k$ is at most $N-1$, the term for $m=N$ in the left-hand side
vanishes and we can as well sum from 0 to $N-1$.
This system is now easily inverted using the sum
$$
\sum_{k=0}^{N-1} (-1)^{k+m} {m\choose k}{k\choose l} = \delta_{l,m},
$$
giving
\bea
y_m &=& \sum_{k=0}^{N-1} (-1)^{k+m} {m\choose k}
(\delta_{k,N-1}-\eta)T^k \nonumber \\
&&\hspace*{-7ex} = (-1)^{N-1+m}{m\choose N-1}T^{N-1} 
-\eta \sum_{k=0}^{N-1} (-1)^{k+m} {m\choose k} T^k, \label{eq:ymsol1}
\nonumber \\
\eea
for $0\le m\le N-1$.
Noting that the first term in (\ref{eq:ymsol1}) is actually zero for
all $m<N-1$, and $T^{N-1}$ for $m=N-1$, and simplifying further, we
find
\bea
y_m &=& -\eta(T-1)^m,\quad 0\le m\le N-2, \label{eq:ymsol2} \\
y_{N-1} &=& T^{N-1}-\eta(T-1)^m.\label{eq:ymsol2n1}
\eea

The value of $y_N$ follows from (\ref{eq:eta}) as
\bea
y_N
&=& T^N\eta - \sum_{m=0}^{N-1} {N\choose m} y_m \nonumber \\
&=& \eta \left[ T^N+ \sum_{m=0}^{N-1} {N\choose m} (T-1)^m \right]
- NT^{N-1} \nonumber \\
&=& \eta(2T^N - (T-1)^N) - NT^{N-1}. \label{eq:yn}
\eea

We can now solve (\ref{eq:ym}) for $x_l$ and $\eta$.
To perform the inversion, we use the sum
$$
\sum_{m=0}^{N-1} (-1)^{k-m} \left(\sum_{p=0}^{N-1} {p\choose
k}{p\choose m}\right) {l+m\choose m} = \delta_{kl},
$$
giving, for $0\le l\le N-1$,
\bea
T^l x_l &=& \sum_{m=0}^{N-1} (-1)^{l-m} \left(\sum_{p=0}^{N-1}
{p\choose l}{p\choose m}\right) \frac{y_m}{(T^2-1)^m} \nonumber \\
&=& -\eta \sum_{m=0}^{N-1} (-1)^{l-m} \left(\sum_{p=0}^{N-1} {p\choose
l}{p\choose m}\right) \frac{1}{(T^2-1)^m} \nonumber \\
&& \times\sum_{k=0}^{N-1} (-1)^{k+m} {m\choose k} T^k \nonumber \\
&& + (-1)^{l-N+1}  {N-1\choose l} \frac{T^{N-1}}{(T^2-1)^{N-1}} \nonumber \\
&=& -\eta (-1)^l\sum_{p=0}^{N-1}{p\choose l}
\left(\frac{T}{T+1}\right)^p \nonumber \\
&& + (-1)^{l-N+1}  {N-1\choose l}
\frac{T^{N-1}}{(T^2-1)^{N-1}}. \label{eq:tx}
\eea

We can solve $\eta$ from (\ref{eq:ym}) and (\ref{eq:yn}), using
(\ref{eq:tx}):
\beas
\lefteqn{\eta \left[2T^N - (T-1)^N \right] - NT^{N-1}} \\
&& \hspace*{-3ex} =(T^2-1)^N \sum_{l=0}^{N-1} {l+N\choose N} T^l x_l \\
&& \hspace*{-3ex} =-\eta(T^2-1)^N \sum_{l=0}^{N-1} {l+N\choose N}
(-1)^l\sum_{p=0}^{N-1}{p\choose l} \left(\frac{T}{T+1}\right)^p \\
&& +  (T^2-1) \sum_{l=0}^{N-1} {l+N\choose N}(-1)^{l-N+1}
{N-1\choose l} T^{N-1} \,.
\eeas
Using the sum
\be
\sum_{l=0}^{N-1} (-1)^l {l+N\choose N}{p\choose l} = (-1)^p {N\choose p}
\ee
this simplifies to
\beas
\lefteqn{\eta \left[2T^N - (T-1)^N\right] - NT^{N-1}} \\
&=& -\eta(T^2-1)^N \left[\left(1-\frac{T}{T+1}\right)^N
-\left(-\frac{T}{T+1}\right)^N\right] \\
&& + (T^2-1)^N N \frac{T^{N-1}}{(T^2-1)^{N-1}} \\
&=& -\eta(T-1)^N (1-(-T)^N) + N(T^2-1)T^{N-1},
\eeas
giving, after solving for $\eta$ and simplifying,
\be\label{eq:etasol}
\eta = \frac{NT}{(2-(1-T)^N)}.
\ee
Inserting this in (\ref{eq:tx}) gives the final solution for
$x_l$. However, we actually need $A_{N-1,l}$, which is given by
$A_{N-1,l} = \det(a)  x_l$.
With the value for $\det(a)$ calculated as (\ref{eq:deta}), this becomes
\bea
\lefteqn{A_{N-1,l}=}\nonumber \\
&& (T^2-1)^{\frac{N(N-1)}{2}} \left[ 2-(1-T)^N \right]
T^{-l} \nonumber \\
&& \times \Big[-\frac{NT}{(2-(1-T)^N)} (-1)^l\sum_{p=0}^{N-1}{p\choose
l} \left(\frac{T}{T+1}\right)^p \nonumber \\
&& + (-1)^{l-N+1}  {N-1\choose l} \frac{T^{N-1}}{(T^2-1)^{N-1}}\Big]
\nonumber \\
&=& N(T^2-1)^{\frac{N(N-1)}{2}}(-1)^{-l+1} T^{-l+1}
\sum_{p=0}^{N-1}{p\choose l} \left(\frac{T}{T+1}\right)^p \nonumber \\
&& + (-1)^{N-l+1}(T^2-1)^{\frac{(N-2)(N-1)}{2}} \left[ 2-(1-T)^N \right]
\nonumber \\ && \times
{N-1\choose l} T^{N-l-1}.\label{eq:sola1}
\nonumber \\
\eea
We now specialise to the case of interest, namely where $T$ takes the
optimal value $T=1-2^{1/N}$.
Then the second term of (\ref{eq:sola1}) vanishes, giving
\bea
\lefteqn{A_{N-1,l}=} \nonumber \\
&& N(T^2-1)^{\frac{N(N-1)}{2}}(-1)^{-l+1} T^{-l+1}
\sum_{p=0}^{N-1}{p\choose l}
\left(\frac{T}{T+1}\right)^p.\label{eq:sola2} \nonumber \\
\eea
We have not substituted the optimal value for $T$ in every part of the
expression, because no simplifications occur at this point.

We finally have to add up $|A_{N-1,l}|$ over all $l$.
We first show that in the optimal $T$, the $A_{N-1,l}$ alternate in
sign.
The factor that could change sign in $(-1)^{N-l-1}A_{N-1,l}$ is
$$
s_l=T^{-l} \sum_{p=0}^{N-1}{p\choose l} \left(\frac{T}{T+1}\right)^p.
$$
The generating function of this sequence is
$$
S(z)=\sum_{l=0}^{N-1}s_l z^l = \frac{(T+1)^N-(T+z)^N}{(T+1)^{N-1}\,(1-z)},
$$
whence it follows that $s_l$ can be rewritten as
\beas
s_l&=&\frac{1}{(T+1)^{N-1}}\sum_{j=l+1}^N {N\choose j}T^{N-j} \\
&=&\frac{1}{(T+1)^{N-1}}\sum_{t=0}^{N-l-1} {N\choose t}T^t.
\eeas
We consider now the functions $f_{j,N}(T):=\sum_{t=0}^{j} {N\choose t}T^t$
and show that they are non-negative for $T\ge -1/N$, which covers the
case of the optimal $T=1-2^{1/N}$. 
First put $T=z-1/N$, which gives after some simplifications
$$
f_{j,N}(z-1/N) = \sum_{l=0}z^l{N\choose l}
\sum_{t=0}^{j-l}{N-l\choose t}(-1/N)^t. 
$$
This is non-negative if the coefficients
$\sum_{t=0}^{j-l}{N-l\choose t}(-1/N)^t$ are non-negative. 
This in turn would follow from the statement $f_{j-l,N-l}(-1/(N-l))\ge0$.
Now the terms in the sum
$$
f_{j,N}(T)=\sum_{t=0}^{j} {N\choose t}(-1/N)^t
$$
constitute an alternating sequence, with the terms for even $t$
positive and the terms for odd $t$ negative. 
Non-negativity of the sum then follows from the fact that the terms
decrease in absolute value: 
indeed, the absolute value of the $t$-term divided by the $t-1$ term
is $(N-t+1)/Nt$, which does not exceed 1 for $t>0$. 
This finally shows non-negativity of the sequence $s_l$.

As a consequence, we can now just add up all $(-1)^{N-l-1}A_{N-1,l}$
and take the absolute value afterwards. This gives
\beas
\lefteqn{\sum_{l=0}^{N-1} (-1)^{N-l-1} A_{N-1,l}} \nonumber \\
&=& N(T^2-1)^{\frac{N(N-1)}{2}}(-1)^N T \sum_{l=0}^{N-1}
\sum_{p=0}^{N-1}{p\choose l} T^{-l}\left(\frac{T}{T+1}\right)^p \\
&=& N(T^2-1)^{\frac{N(N-1)}{2}}(-1)^N T \sum_{p=0}^{N-1}(1+1/T)^p
\left(\frac{T}{T+1}\right)^p \\
&=& N^2 T(T^2-1)^{\frac{N(N-1)}{2}}(-1)^N.
\eeas

The final result is then
\be
\label{eq:den}
\left( \sum\limits_{m=1}^N |A_{Nm}|\right)^2 = N^4 T^2 (T^2-1)^{N(N-1)} \,.
\ee
Combining the results from Eqs.~(\ref{eq:num}) and (\ref{eq:den})
and inserting into the expression (\ref{eq:prob}) for the maximal
success probability we obtain
\begin{equation}
p_{\text{max}} = \frac{1}{N^2}\,,
\end{equation}
just as we set out to show.
This means that there exists a beam
splitter network with an $N$-dimensional ancilla state containing
photon numbers $n_l=0\ldots N-1$ such that a generalized nonlinear
sign shift can be performed on an $N+1$-dimensional signal state with
a probability of $1/N^2$.

\section{Conclusions and discussion}
\label{sec:conclusions}

In this article we have shown how the abstract view on linear-optical
networks can be used to derive scaling laws for success
probabilities. Thus far, we have limited ourselves to single-shot
gates without conditional feed-forward dynamics which could in
principle be incorporated by concatenating several of those abstract
networks. We found that the maximal probability of success of
conditionally generating a (single-shot) generalized nonlinear sign
shift gate on $N+1$-dimensional signal states using $N$-dimensional
ancillas with the lowest possible photon numbers $n_l=0\ldots N-1$
scales as $1/N^2$. To our knowledge, this is the first time such
scaling laws have been found. It also hints toward scaling of success
probabilities of certain classes of $N$-qubit gates. This is due to
the fact that multi-qubit quantum gates acting on tensor-product
states with constant photon number can be decomposed into a multi-mode
Mach--Zehnder interferometer-type setup where single-mode conditional
gates are inserted into the interferometer's paths
\cite{Scheel03}. Note, however, that this is not necessarily the
optimal way to implement such gates. For example, a
controlled-$\sigma_z$ gate would work in $1/16$ of all cases only,
whereas a more general network has been found in \cite{Knill02} that
works with a slightly higher probability of success of $2/27$. In
order to find proper upper bounds on such networks, the existing
abstract network would have to be modified by replacing the single
`active' beam splitter by a U($2M$)-network, $M$ being the number
of modes to be acted upon.

We have shown that computing the success probability within the
framework of the abstract network reduces to the calculation of
various Vandermonde-type determinants. To do so, we defined a class of
polynomials related to the Jacobi polynomials. These
$S$-polynomials obey a three-term recursion relation which is given in
Appendix~\ref{sec:appa}, and the elementary symmetric polynomials
associated with them are derived in Appendix~\ref{sec:appb}.

We have restricted our attention to the experimentally most accessible
case in which the ancilla state contains the lowest possible photon
numbers $n_l=0\ldots N-1$. This choice represents the only
analytically solvable case thus far but, on the other hand, is
motivated by the fact that low photon numbers also means low
decoherence which is desirable for possible applications in quantum
information processing. However, the theory presented above is in
principle valid for any admissible choice of ancilla states. It should
be added that the proof technique for deriving upper bounds by
considering dual convex optimization problems \cite{Eisert} can
similarly applied to the situation considered in this article.

\acknowledgments
The first author thanks the UK Engineering and Physical Sciences Research
Council (EPSRC) for support. Numerous discussions with J.~Eisert and
N.~L\"utkenhaus are gratefully acknowledged.
The work of the second author is part of the QIP-IRC
(\texttt{www.qipirc.org}), supported by
EPSRC (GR/S82176/0), and is also supported by the Leverhulme Trust
grant F/07 058/U.
The authors also warmly
thank an anonymous referee for pointing out a simplification to the
derivation of the maximal success probability.
\appendix
\section{Some properties of the $S$-polynomials}
\label{sec:appa}

In this appendix, we briefly summarize some properties of the
$S$-polynomials.
We recall again their definition (\ref{eq:spoly1}):
$$
S_j^{(x)}(l) = P_j^{(0,l-j)}(2x^2-1),
$$
in terms of the Jacobi polynomials
\beas
\lefteqn{P_n^{(\alpha,\beta)}(x)} \\
&=& 2^{-n}\sum_{m=0}^n {n+\alpha \choose m}{n+\beta \choose
n-m}(x-1)^{n-m} (x+1)^m.
\eeas
From this definition one can
derive a three-term recursion relation as
\begin{eqnarray}
\label{eq:ttrr}
k S_k^{(x)}(n) &=& \left[ (x^2-1)(n+k)+2k-1 \right] S_{k-1}^{(x)}(n)
\nonumber \\ &&
-(k-1)x^2 S_{k-2}^{(x)}(n) \,.
\end{eqnarray}
Then, according to Favard's Theorem \cite{Favard}, which states that
if $P=(P_n)_{n\ge 0}$ is a polynomial sequence which satisfies
\begin{enumerate}
\item $P_{n+1}(x) = (A_n x+B_n)P_n(x)-C_n P_{n-1}(x)$; $P_0(x)=1$,
$P_{-1}(x):=0$ and
\item $A_nA_{n-1}C_n>0$,
\end{enumerate}
then there exists a positive measure $\mu$ on the real line such that
$P$ is an orthogonal sequence with orthogonality measure $\mu$, the
polynomials $S_k^{(x)}(n)$ do form an orthogonal sequence for all
$x\ne 0,\pm 1$. This can be seen by noting that
$k^2A_nA_{n-1}C_n=(x^2-1)x^2$ in our case. As for the excluded values
of the parameters $x$, we have that $S_k^{(\pm 1)}(n)=1$ and
$S_k^{(0)}(n)={n \choose k}$ which both do certainly not form
orthogonal sequences. Orthogonality is important if one needs to find
an inversion relation between the powers of $n$ and the polynomials
$S_k^{(x)}(n)$.

\section{Symmetric polynomials over the $S$-polynomial basis}
\label{sec:appb}

We now present a number of technical results about the expansion of
binomial coefficients and related polynomials in terms of the
$S$-polynomials.

For any positive integer $N$, for any $l$, and for an integer $p$
satisfying $p\le N$,
\be\label{eq:noq}
(x^2-1)^N{l\choose p} = \sum_{j=0}^N (-1)^{N-j}\, s_{N-j}(x;p;N)\,
S_j^{(x)}(l),
\ee
with coefficients
\be\label{eq:noqcoeff}
s_{N-j}(x;p;N) = {p \choose j} x^{2(p-j)} (1-x^2)^{N-p}
\ee
that are polynomials of degree $N-j$ in $x^2$.
Specifically, for $p=N$ these formulas simplify to
\be\label{eq:noqpn}
(x^2-1)^N{l\choose N} = \sum_{j=0}^N (-1)^{N-j}\, {N \choose j}
x^{2(N-j)}\, S_j^{(x)}(l).
\ee

To prove (\ref{eq:noq}) and (\ref{eq:noqcoeff}), we work out the
right-hand side of (\ref{eq:noq})  with (\ref{eq:noqcoeff}) and
(\ref{eq:spoly1}) plugged in:
$$
\sum_{j=0}^N (-1)^{N-j} {p \choose j} x^{2(p-j)} (1-x^2)^{N-p}
\sum_{r=0}^j {j\choose r}{l+r\choose r}(x^2-1)^r
$$
Since the binomial coefficients take on the value 0 for $j<0$, $j>p$,
$r<0$ and $r>j$, and since we have the condition $p\le N$, we can drop
the bounds on the summation signs and subsequently move them up front, giving:
$$
\sum_{j,r} (-1)^{p-j} {p \choose j}{j \choose r}{l+r \choose r}
(x^2)^{p-j} (x^2-1)^{N-p+r}.
$$
Now we notice ${p\choose j}{j\choose r} = {p\choose r}{p-r\choose
j-r}$, so that we can calculate the sum over $j$ easily, using
$$
\sum_j {p-r\choose j-r}(-x^2)^{p-j} = (1-x^2)^{p-r}.
$$
It remains to calculate
$$
\sum_r {p\choose r}{l+r \choose r}(1-x^2)^{p-r} (x^2-1)^{N-p+r},
$$
which is nothing but
$$
(x^2-1)^N \sum_r {p\choose r}{l+r \choose r}(-1)^{p-r}.
$$
It now remains to show that the sum in this expression reduces to
${l\choose p}$.
To do that, we note that ${l \choose p}$ is a polynomial of degree $p$
in $l$, with zeroes $0,1,\ldots,p-1$
and leading term $l^p/p!$. We will confirm that the same holds for the
sum $\sum_r {p\choose r}{l+r \choose r}(-1)^{p-r}$.
Consider first the specific values $l=0,1,\ldots,p-1$.
The factor ${l+r\choose r}$ is a degree $l$ polynomial in $r$. It can
therefore be written as a linear combination
of the degree $j$ polynomials ${r\choose j}$. Thus, with $c_j$
independent of $r$,
\beas
\lefteqn{\sum_{r=0}^p {p\choose r}{l+r\choose r}(-1)^{p-r}} \\
&=& \sum_{j=0}^l c_j(l) \sum_{r=0}^p {p\choose r}{r\choose j}(-1)^{p-r} \\
&=& \sum_{j=0}^l c_j(l) {p\choose j} \sum_{r=0}^p {p-j\choose r-j}(-1)^{p-r}.
\eeas
The sum over $r$ in the latter expression is clearly 0 for $j<p$,
hence the whole expression is indeed zero for $l=0,1,\ldots,p-1$.
Now, concerning the sum's leading term in $l$, we note that it can
only come from the term with summation index $r=p$,
${l+p\choose p}$, whose leading term in $l$ is $l^p/p!$, as
required. This completes the proof of (\ref{eq:noqcoeff}).

The binomial coefficient ${l\choose p}$ is a polynomial in $l$:
$$
{l\choose p} = \frac{1}{p!}l(l-1)\ldots(l-p+1) =
\frac{1}{p!}\prod_{k=0}^{p-1}(l-k).
$$
We now look at related polynomials where one or two of the factors
$(l-k)$ are missing.

Consider first the polynomial
$$
\frac{{l\choose p}}{l-q} = \frac{1}{p!}\prod_{k=0\atop k\neq q}^{p-1}(l-k),
$$
where $q$ is an integer in the range $0\le q< p$.
This polynomial can be expanded in terms of the
binomial coefficients ${l\choose r}$, with $r<p$:
$$
\frac{{l\choose p}}{l-q} = \frac{1}{p{p-1\choose q}} \sum_{r=0}^{p-1}
(-1)^{p-1-r} {r\choose q} {l\choose r}.
$$

The proof of this expansion goes as follows.
Consider the numerator:
$$
\sum_{r=0}^{p-1} (-1)^{p-1-r} {r\choose q} {l\choose r}.
$$
This can again be rewritten using
\begin{eqnarray*}
\lefteqn{
\sum_{r=0}^{p-1} (-1)^{p-1-r} {l-q\choose r-q} } \\
&&=\sum_{s=0}^{p-q-1} (-1)^{p-q-1-s} {l-q\choose s} \\
&&={l-q-1\choose p-q-1}.
\end{eqnarray*}
Wrapping up everything yields the expression
$$
\frac{{l\choose q}{l-q-1\choose p-q-1}}{p{p-1\choose q}},
$$
which is very easily seen to be identical to ${l\choose p}/(l-q)$.

Using this expansion and (\ref{eq:noqpn}), one can easily show the
following. For any positive integer $N$, for any $l$, and for an
integer $q$ satisfying $0\le q\le N$,
\be\label{eq:oneq}
(x^2-1)^N\frac{{l\choose N+1}}{l-q} = \sum_{j=0}^N (-1)^{N-j}
s_{N-j}^{(q)}(x;N) S_j^{(x)}(l),
\ee
with coefficients
\bea
\lefteqn{s_{N-j}^{(q)}(x;N)} \nonumber \\
&=& \frac{1}{(N+1){N\choose q}} \sum_{r=0}^N (-1)^{N-r} {r\choose q}
s_{N-j}(x;r;N) \nonumber \\
&=& \frac{1}{(N+1){N\choose q}} \sum_{r=0}^N (-1)^{N-r} {r\choose q}
{r \choose j} x^{2(r-j)} (1-x^2)^{N-r}. \nonumber \\
&& \label{eq:oneqcoeff}
\eea

\section{Simple Vandermondians with gaps in their arguments}
\label{sec:appc}

We have seen that we need to calculate simple Vandermondians of degree
$N-1$ with integer arguments $[0\ldots N-1]\backslash\{k\}$
[see Eq.~(\ref{eq:onegap})].
Because of Eq.~(\ref{eq:detsn}) it is
enough to concentrate on Vandermondians over the power basis.
First, note that the Vandermondian $V_N(x_1,\ldots,x_N)$ can be
decomposed as
\begin{displaymath}
V_N(x_1,\ldots,x_N) = \prod\limits_{i=1}^{N-1}(x_N-x_i)
V_{N-1}(x_1,\ldots,x_{N-1}) \,.
\end{displaymath}
Hence, we can write
\begin{equation}
V_{N-1}^{(s-1)} = \frac{V_N}{\prod_{m\ne s}|s-m|} \,.
\end{equation}
The product in the denominator can be computed as
\begin{eqnarray}
\lefteqn{
\prod\limits_{m=1,m\ne s}^N |s-m| =
\prod\limits_{m=1}^{s-1} (s-m) \prod\limits_{m=s+1}^N (m-s) }
\nonumber \\
&& = \prod\limits_{q=1}^{s-1} q \prod\limits_{p=1}^{N-s} p =
(s-1)!(N-s)! \,.
\end{eqnarray}
Noting that $V_N=\prod\limits_{n=1}^N(n-1)!=V_{N-1}(N-1)!$, we
obtain
\begin{equation}
V_{N-1}^{(s-1)} = V_{N-1} {N-1 \choose s-1} \,.
\end{equation}
With $k=s-1$ and the additional factor $(|T|^2-1)^{(N-1)(N-2)/2}$,
this is just Eq.~(\ref{eq:onegap}).



\end{document}